\documentclass[aps,prl,twocolumn]{revtex4}
\usepackage{graphicx,graphics}
\usepackage{amssymb}
\usepackage{amsmath}

\begin{document}

\title{Double-stage continuous-discontinuous superconducting phase
transition in the Pauli paramagnetic limit of a 3D superconductor: the URu$%
_{2}$Si$_{2}$ case }
\author{V. Zhuravlev and T. Maniv }
\affiliation{Schulich Faculty of Chemistry, Technion-Israel Institute of Technology, Haifa 32000, Israel}
\date{\today }

\begin{abstract}
The sharp suppression of the de-Haas van-Alphen oscillations observed in the
mixed superconducting (SC) state of the heavy fermion compound URu$_{2}$Si$%
_{2}$ is shown to confirm a theoretical prediction of a narrow double-stage
SC phase transition, smeared by fluctuations, in a 3D
paramagnetically-limitted superconductor. The predicted scenario of a second
order transition to a nonuniform (FFLO) state followed by a first order
transition to a uniform SC state, obtained by using a non-perturbative
approach, is also found to be consistent with recent thermal conductivity
measurements performed on this material.
\end{abstract}

\maketitle

The competition between orbital and spin pair-breaking in strongly type-II
superconductors in the Pauli paramagnetic limit is known to control the
occurrence of discontinuous SC transitions \cite{Sarma63,Maki64} at
sufficiently low temperatures and high magnetic fields. It was found
recently, using perturbation expansion in the SC order parameter \cite{MZ08}%
, that in a clean 3D system, the normal-to-SC phase transitions at low
temperatures are of second order, with a SC phase spatially modulated along
the field direction \cite{FF64,LO64}, whereas the transition line from
nonuniform-to-uniform SC\ state was found to be of the first order. This
conclusion was reached, however, on the basis of perturbation theory, which
might not be valid under the present circumstances due to the following
reasons: (1) The jump of the SC order parameter to a finite value at the
first order phase transition, and (2) the oscillatory dependence of the
quartic and higher order terms in the expansion on the modulation wave
number, which makes the utilization of a uniquely defined expression for the
SC free energy meaningless within perturbation theory.

The heavy fermion superconductor URu$_{2}$Si$_{2}$, whose Fermi surface (FS)
may be characterized as 3D \cite{Ohkuni99}, possesses characteristic FS
parameters which favor strong spin pair breaking. In this material a sharp
rise of the thermal conductivity with the decreasing magnetic field just
below $H_{c2}$ at low temperatures was reported very recently \cite%
{Kasahara07}, indicating the existence of a jump in its electronic entropy
associated with a first-order phase transition. Furthermore, earlier
magneto-oscillations measurements on this material \cite{Ohkuni99} revealed
a very sharp damping of the de Haas-van Alpen (dHvA) effect just below $%
H_{c2}$ which seems to correlate with the anomaly observed in the thermal
conductivity.

In this communication we present results of a non-perturbative approach,
which establishes the sharp, double-stage transition picture, conjectures in
Ref.\cite{MZ08}, and argue by means of a detailed theoretical analysis of
the experimental dHvA data, that the predicted double-stage transition is
realized in URu$_{2}$Si$_{2}$. The proposed model is also shown to be
consistent with the anomaly in the thermal conductivity data reported in Ref.%
\cite{Kasahara07}.

We start by writing an expansion of the thermodynamical potential (TP), $%
\Omega $, in the SC order parameter, $\Delta \left( \mathbf{r}\right) $,
using BCS theory for an isotropic 3D electron gas with the usual $s$-wave
electron pairing, as presented in \cite{MZ08}. \ The use of conventional
pairing was made for the sake of simplicity. \ This is justified in the
clean limit considered here since the relevant results have shown in Ref.%
\cite{MZ08} to be independent of the type of electron pairing.\ Thus we
write: 
\begin{eqnarray}
\Omega \left( \Delta _{0}\right) &=&V\frac{\Delta _{0}^{2}}{g_{int}}%
+\sum_{n=1}\frac{\left( -1\right) ^{n}}{n}\Omega _{2n}\left( \Delta
_{0}\right) ,  \label{ThPotGen} \\
\Omega _{2n} &=&\int d^{3}\left\{ \mathbf{r}\right\} \widetilde{\Gamma }%
_{2n}(\left\{ \mathbf{r}\right\} ,\Delta _{0})\widetilde{K}_{2n}(\left\{ 
\mathbf{r}\right\} )  \notag
\end{eqnarray}%
where $g_{int}$ is the effective BCS coupling constant, $V$ is the volume: 
\begin{eqnarray}
&&\widetilde{\Gamma }_{2n}(\left\{ \mathbf{r}\right\} ,\Delta _{0})=g^{\ast
}(\mathbf{r}_{1},\mathbf{r}_{2})g(\mathbf{r}_{2},\mathbf{r}_{3})...g^{\ast }(%
\mathbf{r}_{2n-1},\mathbf{r}_{2n})  \notag \\
&&\ \ \ \ \ \ \times g(\mathbf{r}_{2n},\mathbf{r}_{1})\Delta (\mathbf{r}%
_{1})\Delta ^{\ast }(\mathbf{r}_{2})...\Delta (\mathbf{r}_{2n-1})\Delta
^{\ast }(\mathbf{r}_{2n})  \label{Vertex}
\end{eqnarray}%
and: 
\begin{eqnarray}
&&\widetilde{K}_{2n}(\left\{ \mathbf{r}\right\} )=k_{B}T\sum_{\nu }\overline{%
G}_{0\downarrow }^{\ast }(\mathbf{r}_{1},\mathbf{r}_{2},\omega _{\nu })%
\overline{G}_{0\uparrow }(\mathbf{r}_{2},\mathbf{r}_{3},\omega _{\nu }) 
\notag \\
&&\ \ \ \ \ ...\overline{G}_{0\downarrow }^{\ast }(\mathbf{r}_{2n-1},\mathbf{%
r}_{2n},\omega _{\nu })\overline{G}_{0\uparrow }(\mathbf{r}_{2n},\mathbf{r}%
_{1},\omega _{\nu })  \label{Kernel}
\end{eqnarray}

Here $\Delta _{0}^{2}=V^{-1}\int d^{3}\mathbf{r}_{i}\left\vert \Delta (%
\mathbf{r}_{i})\right\vert ^{2}$, and $\left\{ \mathbf{r}\right\} \mathbf{=}%
\left\{ \mathbf{r}_{1},...,\mathbf{r}_{2n}\right\} $ denotes the entire set
of position vectors for a cluster consisting of $n$ electron pairs. Note
that for convenience we incorporated the gauge factors, $g(\mathbf{r}_{i},%
\mathbf{r}_{i+1})$, of the Green's functions, $G_{0\uparrow \downarrow }(%
\mathbf{r}_{i},\mathbf{r}_{i+1},\omega _{\nu })$, for a free electron in a
uniform magnetic field, into the vertex part, $\widetilde{\Gamma }_{2n}$, so
that the effective kernel $\widetilde{K}_{2n}$ is given in Eq.(\ref{Kernel})
by a product of the gauge invariant Green's functions, $\overline{G}%
_{0\uparrow \downarrow }(\mathbf{r}_{i},\mathbf{r}_{i+1},\omega _{\nu })$. A
useful expression for such a Green's function for a positive Matsubara
frequency, $\nu \geq 0$, can be written as: 
\begin{eqnarray}
&&\overline{G}_{0\uparrow \downarrow }(\mathbf{r}_{1},\mathbf{r}_{2},\omega
_{\nu })=\frac{1}{2\pi a_{H}^{3}\hbar \omega _{c}}\int \frac{dk_{z}}{2\pi }%
e^{ik_{z}(z_{2}-z_{1})}e^{-\rho _{12}^{2}/4}  \notag \\
&&\int_{0}^{\infty }d\tau e^{i\tau \left[ n_{F}+g-x^{2}+i\varpi _{\nu }%
\right] }\left( 1-e^{-i\tau }\right) ^{-1}\exp \left( -\frac{\rho
_{12}^{2}e^{-i\tau }}{2\left( 1-e^{-i\tau }\right) }\right) ,  \notag
\end{eqnarray}%
where $\mathbf{\rho }_{1,2}\equiv \mathbf{r}_{\bot 2}-\mathbf{r}_{\bot 1}$ ,
with $\mathbf{r}_{\bot 1},\mathbf{r}_{\bot 2}$-the projections of the
initial and final electron position vectors, respectively, on the ($x-y$ )
plane perpendicular to the magnetic field. This expression is obtained after
summation over the Landau level (LL) index $n=0,1,...,$ of the
single-particle energy, $\varepsilon _{n\uparrow \downarrow }/\hbar \omega
_{c}=n+\widetilde{k}_{z}^{2}\mp g-n_{F}+i\varpi _{\nu }$ for a spin up ( or
down ) electron in a magnetic field $\mathbf{H}=H\widehat{z}$ , with a
cyclotron frequency $\omega _{c}=eH/m^{\ast }c$ , Zeeman spin energy $\mp
eH/m_{0}c$, and g-factor $g\equiv m^{\ast }/m_{0}$, with $m^{\ast },$ and $%
m_{0}$- the effective mass and free electron mass respectively. Here $\ 
\widetilde{k}_{z}^{2}=\hbar ^{2}k_{z}^{2}/2m^{\ast }\hbar \omega _{c}$, $%
n_{F}=\mu /\hbar \omega _{c}$, $\mu $-the\textsl{\ }chemical potential ($%
\approx E_{F}$ -Fermi energy), and $\varpi _{\nu }=\omega _{\nu }/\omega
_{c} $,with $\omega _{\nu }=\pi k_{B}T\left( 2\nu +1\right) /\hbar $. \ For
negative Matsubara frequencies, $\nu <0$, a similar expression can be
derived by replacing $\tau $ with $-\tau $. In what follows we will express
space coordinates and momenta in units of $a_{H}=\sqrt{c\hbar /eH}$ and $%
a_{H}^{-1}$ respectively.
The SC order parameter is assumed to take the form, $\Delta (\mathbf{r}%
,z)=\Delta _{\max }e^{iqz}\varphi _{0}(x,y)$, where $\Delta _{\max
}^{2}=\left( \frac{2\pi }{a_{x}^{2}}\right) ^{1/2}\Delta _{0}^{2}$ , and $%
\varphi _{0}(x,y)$ describes (in the symmetric gauge) an hexagonal vortex
lattice with inter-vortex distance, $a_{x}=\frac{\sqrt{2\pi }}{3^{1/4}}$.
Here $e^{iqz}$ is a Fulde-Ferrel (FF) \cite{FF64} modulation function along
the magnetic field direction,controlled by the wave-number, $q$.

The vertex part $\widetilde{\Gamma }_{2n}(\left\{ \mathbf{r}\right\} ,\Delta
_{0})$, Eq.(\ref{Vertex}), is a violently oscillating function of the
lateral relative electronic coordinates, which interferes strongly with the
oscillatory electronic kernel $\widetilde{K}_{2n}(\left\{ \mathbf{r}\right\}
)$, Eq.(\ref{Kernel}). Multiple integration over these coordinates yields
gross cancellations except near stationary configurations, which restrict
all $2n$ electronic position vectors to a relative proximity region of size
of a magnetic length \cite{ZM97},\cite{RMP01}. Other contributions to this
integral, arising from non-stationary, separately paired configurations,
become increasingly important in random vortex lattices where the phase
coherence responsible for the constructively interfering configurations
breakdown\cite{ZM97}. \ Whereas the small oscillatory (high harmonic in $1/H$%
) part of the TP is strongly influenced by these non-local contributions,
their influence on the much larger non-oscillatory (zero harmonic in $1/H$ )
component is not important (see Ref.\cite{RMP01}). The great advantage of
using the local approximation in Eq.(\ref{ThPotGen}) is in its factorization
with respect to the relative coordinates and its apparent independence of
the center of mass coordinates, which enable us rewriting the integrand in
Eq.(\ref{ThPotGen}) as a separable product of effective single electron
Green's functions. The corresponding $n$-th order term, $\Omega _{2n}$, can
be thus written as a 3D integral over the center of mass momentum in an
effective two-particle Green's function, raised to the $n$-th power, by
performing an appropriate Fourier-transformation, namely: $\Omega
_{2n}\left( \Delta _{\max }\right) =V\frac{k_{B}T}{a_{H}^{3}}\frac{a_{x}}{%
\sqrt{2\pi n}}\widetilde{\Delta }_{\max }^{2n}I_{2n}$ , where $\widetilde{%
\Delta }_{\max }^{2}=\frac{\Delta _{\max }^{2}}{\left( \hbar \omega
_{c}\right) ^{2}}$, and: 
\begin{equation}
I_{2n}=\sum_{\nu >0}\int \frac{d^{2}kdk_{z}}{(2\pi )^{3}}
\label{I_2n}  \left[ \Phi _{\nu }\left( \mathbf{k},k_{z}|g,q\right) \Phi
_{\nu }^{\ast }\left( \mathbf{k},k_{z}|-g,-q\right) \right] ^{n}+cc  \notag
\end{equation}
\begin{equation}
\Phi _{\nu }\left( \mathbf{k},k_{z}|g,q\right) =\int_{0}^{\infty }d\tau
e^{-\tau \left[ \varpi _{\nu }-i\xi _{\mathbf{k},k_{z}}\right] -\frac{1}{2}%
\left( 1+i\tau -e^{i\tau }\right) \left\vert \mathbf{k}\right\vert ^{2}} \notag
\end{equation}
with: $\xi _{\mathbf{k},k_{z}}\left( g,q\right) =\frac{1}{2}\left\vert 
\mathbf{k}\right\vert ^{2}+\frac{1}{2}\left( k_{z}-\frac{q}{2}\right)
^{2}-g-n_{F}$ , \ $\mathbf{k}=\widehat{\mathbf{x}}k_{x}+\widehat{\mathbf{y}}%
k_{y}$ . The resulting perturbation series can be easily summed to all
orders, provided the reduction pre-factor $\frac{1}{\sqrt{n}}$ , arising
from the overlap integral of $n$ lowest LL orbitals is represented as a
Gaussian integral: $\frac{2}{\sqrt{\pi }}\int_{0}^{\infty }exp\left(
-nu^{2}\right) du$ . In the quasi-classical limit $\tau \ll 1$, the Gaussian
approximation $e^{-\frac{1}{2}\left( 1+i\tau -e^{i\tau }\right) \left\vert 
\mathbf{k}\right\vert ^{2}}\simeq e^{-\frac{1}{4}\tau ^{2}\left\vert \mathbf{%
k}\right\vert ^{2}}$, accounts for the diamagnetic pair-breaking whereas all
quantum corrections (including quantum magnetic oscillations), which arise
near the lattice points $\tau =2\pi l$ , $l\not=0$ , are neglected.

It is convenient to normalize all energies by $\pi k_{B}T_{c0}$ , where $%
T_{c0}$ is the transition temperature at $H=0$ , so that: $\overline{\Delta }%
_{\max }=\frac{\Delta _{\max }}{\pi k_{B}T_{c0}}$ , $\overline{\mu }=\frac{%
\mu }{\pi k_{B}T_{c0}}$, $\overline{q}^{2}=\frac{\hbar ^{2}q^{2}/2m^{\ast }}{%
\pi k_{B}T_{c0}}$, $\left\vert \overline{\mathbf{k}}\right\vert ^{2}=\frac{%
\hbar ^{2}\left\vert \mathbf{k}\right\vert ^{2}/2m^{\ast }}{\pi k_{B}T_{c0}}$
, $\overline{k}_{z}^{2}=\frac{\hbar ^{2}k_{z}^{2}/2m^{\ast }}{\pi k_{B}T_{c0}%
}$ and $\overline{g}=\frac{g\hbar \omega _{c20}}{\pi k_{B}T_{c0}}$, where $%
\omega _{c20}\equiv eH_{c20}/m^{\ast }c$, and $H_{c20}$ is the theoretical
upper critical field at $T=0$ (in the absence of spin splitting). \
Performing the integration over $\tau $, the resulting expression for TP
can be written in the form: 
\begin{eqnarray}
&&\frac{\Omega }{V\Xi _{0}} =\frac{\overline{\Delta }_{\max }^{2}}{\lambda }-%
\frac{t}{\pi \overline{\mu }^{1/2}}\mathbf{Re}\sum_{\nu >0}  \notag \\
&&\left\langle \int \overline{k}d\overline{k}d\overline{k}_{z}\ln \left(
1+e^{-u^{2}}\overline{\Delta }_{\max }^{2}\Phi _{\nu }^{+}\Phi _{\nu
}^{-}\right) \right\rangle _{u} , \label{ThPot} \\
&&\Phi _{\nu }^{\pm } =\frac{\sqrt{\pi }}{\left( 2\overline{\omega }%
_{c20}b\right) ^{1/2}\overline{k}}e^{\varepsilon _{\nu }^{\pm }\left(
k,k_{z}\right) ^{2}}\mathbf{erfc}\left[\varepsilon _{\nu }^{\pm }\left(
k,k_{z}\right) \right],   \notag\\
&&\varepsilon _{\nu }^{\pm }\left( k,k_{z}\right) =\frac{t\left( 2\nu
+1\right) \pm i\left( \overline{k}^{2}+\left( \overline{k}_{z}\pm \overline{q%
}\right) ^{2}\pm \overline{g}b-\overline{\mu }\right) }{\left( 2\overline{%
\omega }_{c20}b\right) ^{1/2}\overline{k}},    \notag
\end{eqnarray}%
where $b=H/H_{c20}$, $t=T/T_{c0}$ , $\overline{\omega }_{c20}=\hbar \omega
_{c20}/\pi k_{B}T_{c0}$ , $\ \left\langle f\left( u\right) \right\rangle
_{u}\equiv \frac{2}{\sqrt{\pi }}\int_{0}^{\infty }duf\left( u\right) $ , $%
\frac{1}{\lambda }=\sqrt{2\overline{\mu }^{1/2}}\sum_{\nu >0}\frac{1}{2\nu +1%
},$ and $\Xi _{0}=0.137\left( k_{F}\xi \left( 0\right) \right) ^{2}\frac{%
k_{B}T_{c0}}{\xi \left( 0\right) ^{3}}$, with $\xi \left( 0\right)
=0.18\hbar v_{F}/k_{B}T_{c0}$ the zero temperature coherence length.

\begin{figure}[htb]\label{fig:1}
\includegraphics[width=8cm]{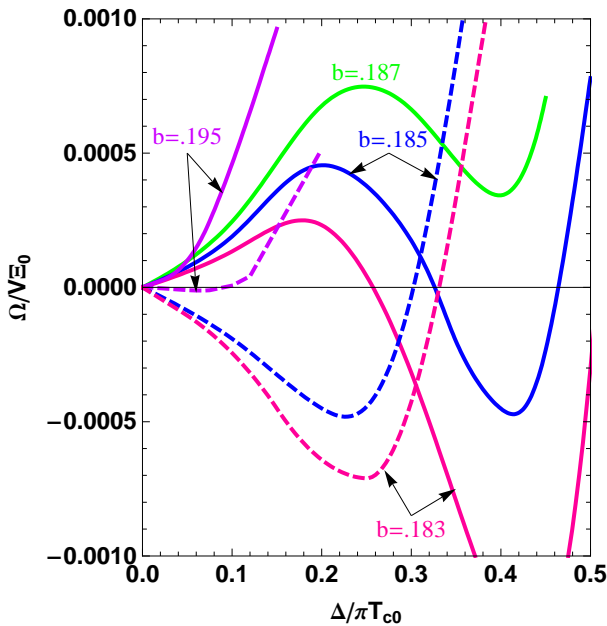}
\caption{(color online) $\Omega $ vs. $\overline{\Delta 
}_{\max }$ (in units of $\Xi _{0}V$) for a uniform ($\ q=0$ ) SC order
parameter (solid curves) and for the corresponding FF modulated order
parameter (dashed curves) at various magnetic field values $b$ near the SC
transition at temperature $t=0.02$. The selected g-factor is $\overline{g}%
=1.8$. Note the second-order transition to the FF state at $b\approx 0.195$
and the first-order transition to the uniform SC state at $b\approx 0.185$.}
\label{fig:1}
\end{figure}

\begin{figure}\label{fig:2}
\includegraphics[width=8cm]{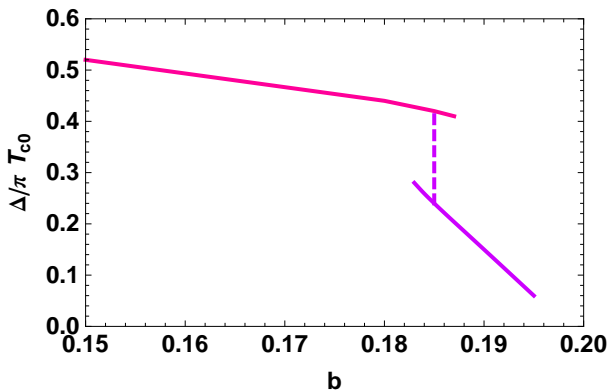}
\caption{(color online) Field dependence of the
self-consistent order-parameter amplitude $\overline{\Delta }_{\max }$ at a
temperature $t=0.02$ , well below the tri-critical point ($t_{tric}\approx
0.4$), for $\overline{g}=1.8$. }
\label{fig:2}
\end{figure}

In the limit of zero spin splitting and in the absence of FF modulation ($%
g=q=0$) the effective pairing parameter, $X_{\nu }\left( k,k_{z}\right)
\equiv \overline{\Delta }_{\max }^{2}\Phi _{\nu }^{+}\Phi _{\nu }^{-}$, is
always real and positive. Under these circumstances the general form of $%
\Omega \left( \overline{\Delta }_{\max }^{2}\right) $ at $H<H_{c20}$, as
expressed in Eq.(\ref{ThPot}), possesses the single minimum structure
characterizing the usual GL theory. \ For $g\neq 0$ ,\ $X_{\nu }\left(
k,k_{z}\right) $ can become complex (a feature that can be "healed" by the
presence of the FF modulation wavenumber $q$ ), so that the general form of $%
\Omega \left( \overline{\Delta }_{\max }^{2}\right) $ may show a maximum at
small $\overline{\Delta }_{\max }^{2}$ which is followed by a minimum at
larger $\overline{\Delta }_{\max }^{2}$. \ The initial maximum reflects the
competition between the increasing spin paramagnetic energy and decreasing
SC pair-correlation energy as the number of spin-singlet Cooper-pairs is
increased.

Typical results of the calculated TP using Eq.(\ref{ThPot}) for non-zero
spin-splitting are show in Fig.\ref{fig:1}.  As discussed above, the restriction to
SC states with $q=0$ , represented in Fig.\ref{fig:1} by the solid curves, leads to
formation of a maximum at small $\overline{\Delta }_{\max }$ and a local
minimum at larger $\overline{\Delta }_{\max }$ by the strong spin splitting
effect as the magnetic field is reduced. Upon further field decrease the
minimum becomes global and so should drive a first order normal-to-SC phase
transition. Allowing for states with $q\neq 0$, however, the unusual feature
(i.e. ${Im\ }X_{\nu }\left( k,k_{z}\right) \neq 0$ ) associated with the
strong spin splitting effect is "healed", and the usual single minimum
picture is restored (see the dashed curves in Fig.\ref{fig:1}). Thus, instead of the
"expected" first-order transition to a uniform SC state one finds a
second-order phase transition to a nonuniform (FF) SC state.

However, due to its compensation effect,
the FF modulation significantly reduces the equilibrium SC free energy with
respect to its uniform counterpart (compare the dashed curves to the
corresponding solid ones). As a result, the field range of stability of the
modulated phase is quite small. Thus, by slightly reducing the field below
the second order normal-to-SC transition the $q=0$ state becomes
energetically more favorable and the system transforms from the nonuniform
to a uniform SC state via a first-order phase transition. One should note
that the second order perturbation theory with $\Omega /V\Xi _{0}=\alpha
\left( q\right) \overline{\Delta }_{\max }^{2}+\frac{1}{2}\beta \left(
q\right) \overline{\Delta }_{\max }^{4}$,\ is quantitatively correct only for
small values of the order parameter,namely for $\overline{\Delta }_{\max
}\lesssim .1$, and, therefore, cannot be applied to the first order
transition.

Fig.\ref{fig:2} shows the calculated self-consistent $\overline{\Delta }_{\max }^{2}$
(which is optimized with respect to $q$ ) as a function of magnetic field at
a temperature well below the tri-critical temperature and for characteristic
parameters corresponding to URu$_{2}$Si$_{2}$. \ The initial build-up of the
SC order parameter in a narrow region following the second-order phase
transition and the pronounced jump at the next (first order) transition are
apparent. The overall width of the two-stage transition ($\Delta b/b\approx
0.06$\ ) is in good agreement with the width of the sharp structure observed
experimentally in the thermal transport measurements \cite{Kasahara07}.

\begin{figure}
\includegraphics[width=7cm]{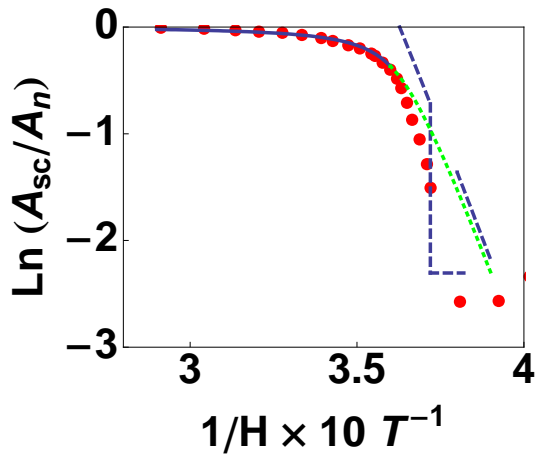}
\caption{(color online)  Logarithm of the dHvA amplitude
ratio $A_{SC}/A_{n}$ reported in Ref.\protect\cite{Ohkuni99} (full circles)
and the values of $-\protect\chi \left\langle \Delta ^{2}\right\rangle ,%
\protect\chi =\frac{\protect\pi ^{3/2}}{\hbar \protect\omega _{c}\protect%
\sqrt{E_{F}\hbar \protect\omega _{c}}}$ as obtained from our calculation
near the second order phase transition (solid line), as functions of $1/H$.
The dashed broken straight line is our mean-field result for $-\protect\chi %
\Delta _{\max }^{2}$\ around the first-order transition. The doted line
represents the extrapolation of $-\protect\chi \left\langle \Delta
^{2}\right\rangle $ for the modulated FF state to the low-field regime. }
\label{fig:3}
\end{figure}

Fig.\ref{fig:3} exhibits the result of a detailed fitting procedure of our calculation
to the experimental data of the dHvA oscillations observed by Ohkuni et. al 
\cite{Ohkuni99} in the mixed state of URu$_{2}$Si$_{2}$. \ The data shows a
very sharp reduction in the amplitude of the dHvA oscillation, $A_{SC}$ ,
just below $H=2.78$ T , similar to the step-like structure observed in the
thermal conductivity measurements \cite{Kasahara07}. For fitting the
measured relative signal, $A_{SC}/A_{n}$ ( $A_{n}$ being the theoretical
dHvA amplitude as extrapolated from the normal state) in the region above
the sharp damping interval, we exploit the fluctuating vortex lattice model
described in Refs.\cite{RMP01},\cite{Maniv06} and write: $\ln \left(
A_{SC}/A_{n}\right) =-\frac{\pi ^{3/2}}{\hbar \omega _{c}\sqrt{E_{F}\hbar
\omega _{c}}}\left\langle \Delta ^{2}\right\rangle $ , where $\left\langle
\Delta ^{2}\right\rangle $ is the mean-square order parameter\ for the
modulated FF state in the vicinity of the second-order phase
transition,i.e.: $\left\langle \Delta ^{2}\right\rangle =\frac{a_{x}}{2\sqrt{%
2\pi }}\Delta _{\max }^{2}\left( 1+\sqrt{1+\nu _{\kappa }^{2}/x^{2}}\right) $%
, with $x=\frac{a_{x}}{\sqrt{2\pi }}\sqrt{\frac{\beta \Xi _{0}V_{0}}{2k_{B}T}%
}\overline{\Delta }_{\max }^{2}$, and $\nu _{\kappa }=0.51$ for a 3D system
(see Ref.\cite{Maniv06}). Here $\beta $ is the coefficient of the quartic
term obtained in the expansion of $\frac{\Omega }{V\Xi _{0}}$ in $\overline{%
\Delta }_{\max }^{2}$ near the SC transition, and $V_{0}=\pi
^{2}a_{H_{c2}}^{2}/k_{F}$. The selected value of $E_{F}$ has been determined
from the experimentally observed, dominant dHvA frequency, $F_{\alpha
}\approx 10^{3}$ T, corresponding to the nearly spherical band 17-hole Fermi
surface reported in Ref.\cite{Ohkuni99}.

The values of $\overline{\Delta }_{\max }^{2}$ at $H<H_{c2}$ were determined in this calculation by
minimizing the TP $\Omega $ with respect to both $\overline{\Delta }_{\max
}^{2}$ and $\overline{q}$ (see Fig.\ref{fig:2}), whereas at $H>H_{c2}$ they were
obtained by the analytical continuation of the perturbative mean field
expression, $\overline{\Delta }_{\max }^{2}=-\alpha /\beta $ into the region 
$-\alpha /\beta <0$, where $\alpha $ and $\beta $ are the quadratic and
quartic coefficients respectively in the perturbation expansion of $\Omega
\left( \overline{\Delta }_{\max }^{2}\right) $. The optimal value of $%
\overline{q}$ in this region was selected by minimizing with respect to $%
\overline{q}$ the free energy obtained from the functional integral of $%
e^{-\Omega \left( \overline{\Delta }_{\max }^{2}\right) /k_{B}T}$ over $%
\overline{\Delta }_{\max }^{2}$, which amounts in the Gaussian approximation
to minimizing $\alpha $ with respect to $\overline{q}$. \ In the fitting
procedure we have exploited the interpolation formula \ $\left\langle \Delta
^{2}\right\rangle =\frac{1}{2}D_{0}^{2}\left( 1-H/H_{c2}\right) +\sqrt{%
\alpha _{I}^{2}+\frac{1}{4}D_{0}^{4}\left( 1-H/H_{c2}\right) ^{2}}$\cite%
{Maniv06} with the best fitting adjustable parameters $D_{0}=4.2\ K$, $%
H_{c2}=2.76\ T$ , and $\alpha _{I}=.06\ K$ (see Fig.\ref{fig:3}). The resulting value
of the order parameter is in a reasonably good agreement with the zero-field
gap parameter $\Delta _{BCS}=1.76k_{B}T_{c0}=2.5\ K$ obtained within BCS
theory with the experimental value of $T_{c0}\ $($=1.4\ K$ ). \ 

Using the resulting parameters and the field dependent $\overline{\Delta }%
_{\max }^{2}$ shown in Fig.\ref{fig:2}, we have calculated the jump of $\ln \left(
A_{SC}/A_{n}\right) $ at the first order phase transition within our
nonperturbative mean field theory. The agreement with the experimental data
is good, keeping in mind that thermal fluctuations and inhomogeneous
broadening could be responsible for the observed smearing of the theoretical
discontinuous transition. Nevertheless, the sharp downward deviation of the
experimental data in Fig.\ref{fig:3} from the extrapolation of the calculated $-\chi
\left\langle \Delta ^{2}\right\rangle $ for the modulated FF state to the
low-field regime provides a strong evidence for the two-stage nature of the
SC transition. It should be stressed that the relative size of the jump in $%
\Delta _{\max }^{2}$ obtained in our calculation is nearly independent of
the various parameters involved, provided the temperature $t$ is well below
the tri-critical point $t_{tric}$.

In conclusion, using a non-perturbative approach, we have firmly established
our early conjecture concerning the SC transition in a 3D strongly type-II
superconductor in the paramagnetic limit, and show that the dHvA effect
observed in the mixed SC state of URu$_{2}$Si$_{2}$\cite{Ohkuni99} provides
a clear experimental evidence for the double-stage nature of this
transition, which is smeared by significant SC fluctuations effect. This
finding is consistent with the interpretation of a first-order phase
transition given in Ref.\cite{Kasahara07} to the step-like structure
observed in the thermal transport data of this material. We note that the
unusual sign of the observed jump in the thermal conductivity, which could
be due to some peculiar quasi-particle scattering mechanism \cite%
{AdachiSigrist08}, is irrelevant to our main argument, which associates this
jump, irrespective of its direction, to the jump of the SC order parameter
at the predicted first-order transition.

We thank J. Wosnitza, B. Bergk and Y. Kasahara for valuable discussions.
This research was supported by the Israel Science Foundation founded by the
Academy of Sciences and Humanities, by Posnansky Research fund in
superconductivity, and by EuroMagNET under the EU contract
RII3-CT-2004-506239.


\begin{thebibliography}{99}
\bibitem{Sarma63} G. Sarma, J. Phys. Chem. Solids \textbf{24 }, 1029 (1963).

\bibitem{Maki64} K. Maki and T. Tsuneto, Prog. Theor. Phys. \textbf{31}, 945
(1964).

\bibitem{MZ08} T. Maniv and V. Zhuravlev, Phys. Rev. B \textbf{77} , 134511
(2008).

\bibitem{FF64} P. Fulde and R.A. Ferrell, Phys. Rev. \textbf{135}, A550,
(1964).

\bibitem{LO64} A.I. Larkin and Yu.N. Ovchinnikov, Zh. Eksp. Teor. Fiz. 
\textbf{47}, 1136 (1964) [Sov. Phys. JETP \textbf{20}, 762 (1965)]

\bibitem{Ohkuni99} H. Ohkuni, Y. Inada, Y. Tokiwa, K. Sakurai, R. Settai, T.
Honma, Y. Haga, E. Yamamoto. Y. Onuki, H. Yamagami, S. Takahashi and T.
Yanagisawa, Phil. Mag. B \textbf{79}, 1045 (1999).

\bibitem{Kasahara07} Y. Kasahara et al., Phys. Rev. Lett. \textbf{99},
116402 (2007).

\bibitem{ZM97} V. Zhuravlev, T. Maniv, I. D. Vagner, and P. Wyder, Phys.
Rev. B \textbf{56} , 14693 (1997).

\bibitem{RMP01} T. Maniv, V. Zhuravlev, I. D. Vagner, and P. Wyder, Rev.
Mod. Phys.,\textbf{73}, 867.

\bibitem{Maniv06} T. Maniv, V. Zhuravlev, J. Wosnitza, O. Ignatchik, B.
Bergk, and P.C. Canfield, Phys. Rev. B \textbf{73}, 134521-7 (2006).

\bibitem{AdachiSigrist08} Hiroto Adachi and M. Sigrist, arXiv
[cond-mat.super-con]: 0710.3110
\end{thebibliography}
\end{document}